\documentclass{iau}
\usepackage{graphicx}

\title[A new code coupling collisions and dynamics in debris discs] %% give here short title %%
{LIDT-DD: A new self-consistent debris disc model including
radiation pressure and coupling dynamical and collisional
evolution}

\author[Q. Kral, P. Thebault, S. Charnoz]   %% give here short author list %%
{Q. Kral$^1$,
 P. Thebault$^1$, S. Charnoz$^2$}

\affiliation{$^1$LESIA-Observatoire de Paris, UPMC Univ. Paris 06, Univ. Paris-Diderot, France \\ email: {\tt quentin.kral@obspm.fr} \\[\affilskip]
$^2$Laboratoire AIM, Universit«e Paris Diderot / CEA / CNRS, Institut Universitaire de France}

\pubyear{2013}
\volume{299}  %% insert here IAU Symposium No.
\pagerange{119--126}
\jname{Exploring the Formation and Evolution of Planetary Systems}
\editors{Brenda Matthews, James Graham}
\begin{document}

\maketitle

\begin{abstract}
The first attempt at developing a fully self-consistent
code coupling dynamics and collisions to study debris discs (Kral,
Thebault, Charnoz, 2013) is presented. So far, these two crucial mechanisms were
studied separately, with N-body and statistical collisional codes
respectively, because of stringent computational constraints. In
particular, incorporating collisional effects (especially destructive
collisions) into an N-body scheme was deemed an impossible task because of
the exponential increase of particles it would imply.

We present here a new model named LIDT-DD which is able to follow
over long timescales the coupled evolution of dynamics (including radiation pressure)
and collisions in a self-consistent way.

%In this 3D Lagrangian-Eulerian code,
%grains of a given size at a given location in a disc are grouped into
%"super-particles" (SPs), whose orbits are tracked with an N-body code and
%whose mutual collisions are treated using a particle-in-a-box scheme. To
%handle the complexity of grain dynamics, a sorting procedure, regrouping
%all SPs into dynamical families, has been implemented. A complex SP
%reassignment routine, looking for and reallocating all redundant SPs,
%prevents their number from diverging.

%Our code has been tested for a set of simplified cases for which we
%reproduce well-known robust results. LIDT-DD has also been used on
%a first test case: the violent breakup of a massive planetesimal within a
%debris disc. I will present some preliminary results for this case, for
%which we are, for the first time, able to quantify the survival time of
%the signatures left by such violent transient events. I will conclude by
%showing a list of all potential applications of this code to debris disc
%studies.
\end{abstract}

\firstsection % if your document starts with a section,
\section{Context \& Objectives}

Up to now, the evolution of debris discs was studied with two radically different approaches: 1) Dynamics (with N-body codes) to study the structures and 2) Collisions (with particle-in-a-box codes) to study the grinding process in such a disc. While such separate studies can (and have) produce(d) important results, they suffer from unavoidable limitations. The absence of collisions in N-body codes can indeed strongly bias or even invalidate some results obtained in such approach: As an example, if collisional timescales are shorter than dynamical ones then collisions can hinder or even prevent the build-up of dynamical structures. For the purely statistical models these limitations are obviously the absence of or poor spatial resolution, but also the fact that dynamical processes might strongly affect impact rates and velocities and thus the collisional evolution.

%Likewise, the identification of dynamically stable and unstable regions in a perturbed debris disc (be it by a planet or a stellar companion) can also be strongly affected by collisional activity, as a collision cascade will steadily produce small grains that can be launched by radiation pressure on highly eccentric or unbound orbits that can populate regions that are in principle dynamically "forbidden". 

The first attempts at partially coupling dynamics and collisions have been published recently (e.g., CGA and DyCoSS codes by \cite{star09} and \cite{theb12b}), most of them taking as a basis the N-body approach into which some collision-imposed properties are injected. But these codes are limited to steady state cases with only one perturber and no fragmentation.

Our new code LIDT-DD overcomes all these limitations and enables coupling dynamics (with radiation pressure) and collisions in a self-consistent way, and is able to follow complex interactions in discs over long timescales.

\begin{figure*}

%\makebox[\textwidth]{
%\includegraphics[scale=0.2]{animmore_rthetaexpltaugoutnid4binkeep001.jpg}
%\includegraphics[scale=0.2]{animmore_rthetaexpltaugoutnid4binkeep009.jpg}
%}
\makebox[\textwidth]{
\includegraphics[scale=0.16]{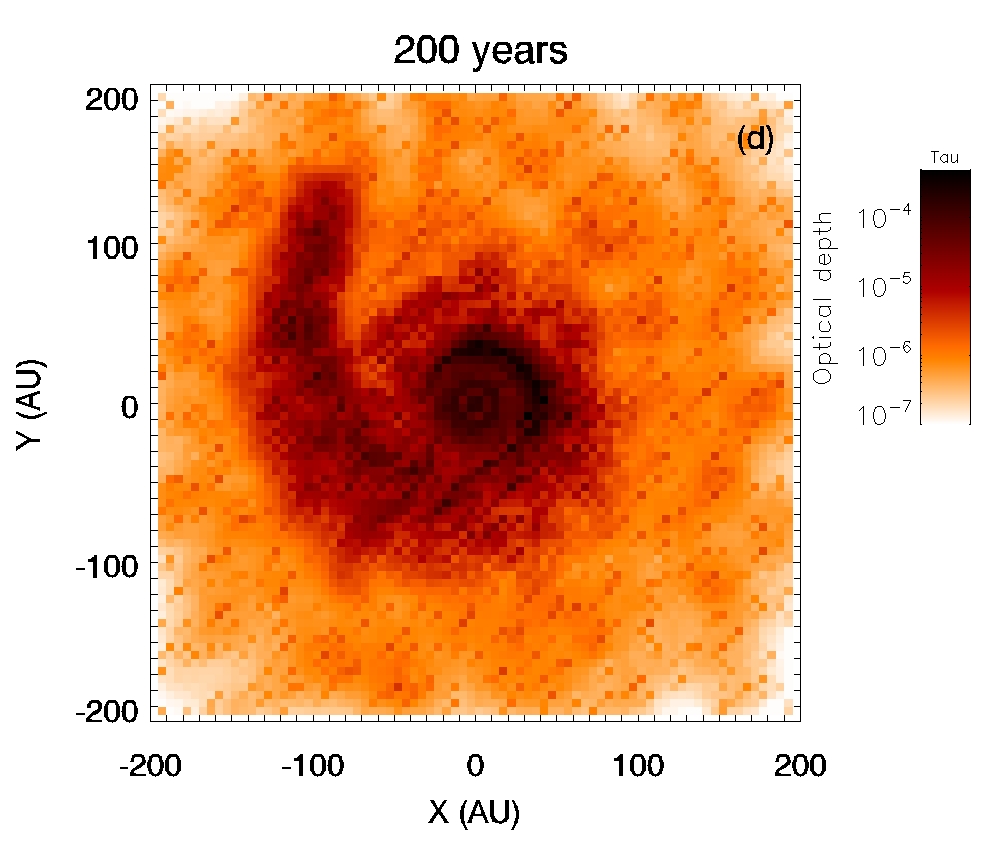}
\includegraphics[scale=0.16]{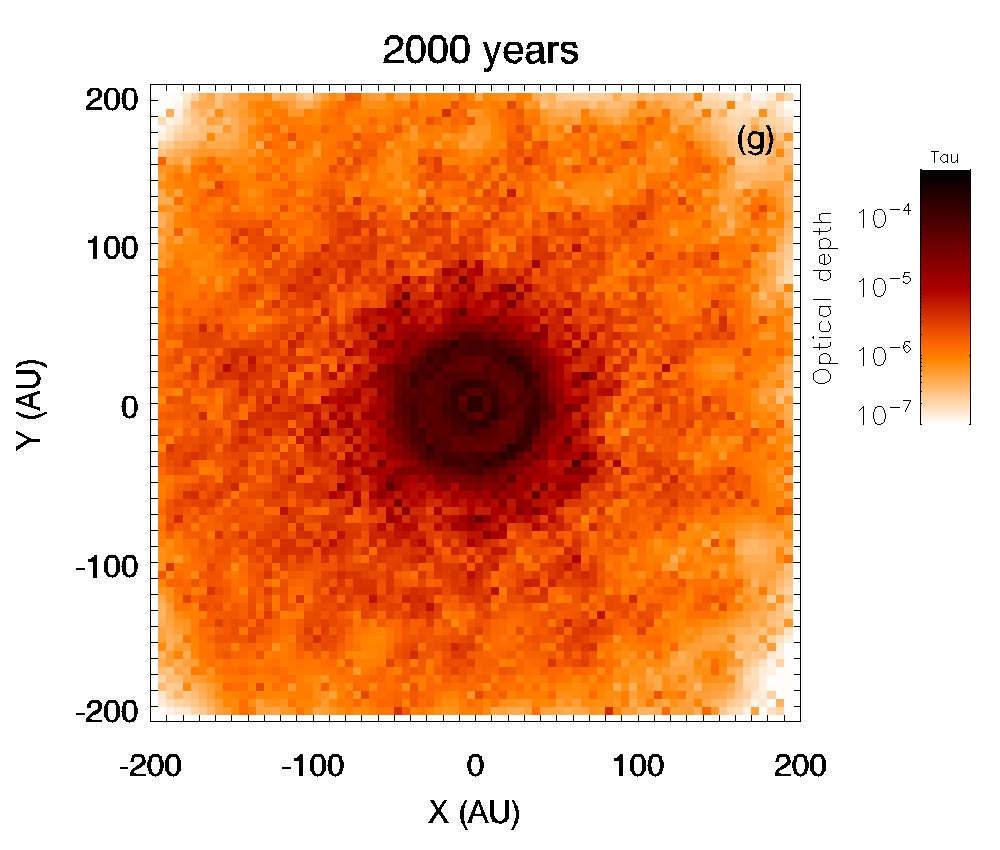}
}
\caption[]{Total smoothed vertical geometrical optical depth evolution in the case of a massive planetesimal breakup within a debris disc.}

\label{optdepthrun}
\end{figure*}

\section{The LIDT-DD code: Principle \& First Results...}
 
LIDT-DD is based on the LIDT3D code
developed by Charnoz et al. (2012) for protoplanetary discs, and strongly
upgraded to account for the complexity of debris disc physics (high
velocity collisions, radiation-pressure affected orbits, wide range of
grains' dynamical behaviour, etc). In this 3D Lagrangian-Eulerian scheme,
grains of a given size at a given location in a disc are grouped into
"super-particles" (SPs), whose orbits are tracked with an N-body code and
whose mutual collisions are treated using a particle-in-a-box scheme. To
handle the complexity of grain dynamics, a sorting procedure, regrouping
all SPs into dynamical families, has been implemented. A complex SP
reassignment routine, looking for and reallocating all redundant SPs,
prevents their number from diverging.

Our code has been tested for a set of simplified cases for which we
reproduce well-known robust results. LIDT-DD has also been used on
a first astrophysical case: the violent breakup of a massive planetesimal within a
debris disc (see Fig~\ref{optdepthrun}). Dust due to the breakup is released within a debris 
disc and its dynamical and collisional evolution is followed. A spiral pattern
develops and fades away in 2000 years. A secondary ring is created because
of keplerian shear and survives for millions of years.

\section{Perspectives}

We intend to use LIDT-DD to explore individual systems with "abnormal" flux excesses, such as HD172555 or TYC 8241 2652 1, and also to investigate how generic the massive-break-up scenario can be for explaining \emph{all} bright debris discs with luminosities that cannot be explained by classical collisional cascades. Another area of interest for LIDT-DD are the planet-disc interactions and the extent to which planetary companions can sculpt debris discs. LIDT-DD enables the exploration of transient events, multi-planet systems as well as the feedback of the planetary perturbations on the collisional evolution. Another potential application of the code is the puzzling case of bright exozodiacal discs, which have been identified by interferometry very close to several main sequence stars. Scenarios for explaining exozodis, such as transient massive impacts, planet scattering, falling evaporating bodies or pile up due to the complex interplay of PR drag, sublimation and collisions, can in principle also be investigated with LIDT-DD.
%More generally, LIDT-DD enables the handling of all complex scenarios where both dynamics and collisions are expected to play an important role in a debris disc evolution.

\end{document}